\documentclass[12pt]{article}
\usepackage{bbm,latexsym,amsmath,epsfig}
\newcommand{\lesssim}{ \ \mbox{\raisebox{-3pt}{$\stackrel%
{\displaystyle <}{\sim}$}} \ }
\newcommand{\gtrsim}{\:\mbox{\raisebox{-3pt}{$\stackrel%
{\displaystyle >}{\sim}$}}\:}
\newcommand{\ten}{\mathbf{10}}
\newcommand{\oht}{\mathbf{120}}
\newcommand{\hts}{\overline{\mathbf{126}}}
\newcommand{\st}{\mathbf{16}}

\newcommand{\mnu}{\mathcal{M}_\nu}
\newcommand{\deltasol}{\Delta m^2_\odot}
\newcommand{\deltaatm}{\Delta m^2_\mathrm{atm}}
\newcommand{\berr}{\!\begin{array}{l} \scriptstyle +}
\newcommand{\tr}{\\[-3mm] \scriptstyle -}
\newcommand{\eerr}{\end{array}}
\newcommand{\ones}%
{\mathbf{1}\oplus\mathbf{1}^{\prime}\oplus\mathbf{1}^{\prime\prime}}
\newcommand{\meff}{|\langle m_{\beta\beta}\rangle|}

\begin{document}

\title{\normalsize \hfill UWThPh-2007-21 \\[1cm]
\LARGE
Embedding the Zee--Wolfenstein Neutrino Mass Matrix in an
$SO(10)\times A_4$ GUT Scenario\\[8mm]}

\author{
Walter Grimus\thanks{E-mail: walter.grimus@univie.ac.at}\,
\normalsize and \large
\setcounter{footnote}{3}
Helmut K\"uhb\"ock\thanks{E-mail: helmut.kuehboeck@gmx.at}
\\
\small Fakult\"at f\"ur Physik, Universit\"at Wien \\
\small Boltzmanngasse 5, A--1090 Wien, Austria
\\*[4.6mm]}

\date{January 22, 2008}

\maketitle

\begin{abstract}
We consider renormalizable $SO(10)$ Yukawa interactions and put the
three fermi\-onic 16-plets into the 3-dimensional irreducible $A_4$
representation. Scanning the possible $A_4$ representation assignments
to the scalars, we find a unique case which allows to accommodate
the down-quark and charged-lepton masses.
Assuming type~II seesaw dominance, we obtain a viable scenario with the
Zee--Wolfenstein neutrino mass matrix, i.e., the Majorana mass matrix
with a vanishing diagonal. Contributions from the charged-lepton mass
matrix resolve the well-known problems with lepton mixing arising from
the vanishing diagonal.
In our scenario, fermion masses and mixings are well reproduced for
both normal and inverted neutrino mass spectra, and $b$--$\tau$ Yukawa
unification and definite predictions for the effective mass in
neutrinoless double-$\beta$ decay are obtained.
\end{abstract}

\newpage

\section{Introduction}
\label{sec-intro}

Grand Unified Theories (GUTs) based on the gauge group
$SO(10)$~\cite{fritzsch} are a
framework for attempts to understand the observed
fermion masses and mixings.
These theories feature a 16-dimensional irreducible
representation (irrep), the spinor representation,
which naturally accommodates all chiral
fermions of one Standard Model (SM) generation
plus a right-handed neutrino.
Furthermore, $SO(10)$ GUTs allow for
type~I~\cite{seesaw} and type~II~\cite{typeII}
seesaw mechanisms (see also~\cite{seesaw-general})
for explaining the smallness of the light neutrino masses.

By employing only one scalar in the $\ten$
and one in the $\hts$ irrep of $SO(10)$
in renormalizable Yukawa couplings,
the so-called
``Minimal Supersymmetric $SO(10)$ GUT'' (MSGUT)~\cite{MSGUT}
has been successful in accounting for all
fermion masses and mixings,
if one focuses solely on its fermion mass
matrices. Moreover,
this model has built-in the gauge-coupling unification
of the Minimal Supersymmetric Standard Model~(MSSM).
In-depth studies of the MSGUT have been
performed~\cite{detailed,schwetz},
also with inclusion of small~\cite{small120}
and prominent~\cite{aulakh06}
effects of the $\oht$ irrep.
Despite its success in reproducing the known fermion
masses and mixings, it should be stressed, however, that the MSGUT
considered as a whole is too constrained, as its scalar sector does
not admit the vacuum expectation values (VEVs) required for the fit
in the fermionic sector---see~\cite{MSGUTout} and~\cite{schwetz}.

However, one is not really satisfied by just reproducing known fermion
masses and mixings, one would also like to explain, for instance, the
threefold replication of fermion generations,
or the peculiar mixing properties of the lepton sector \cite{nu-review},
that is, maximal atmospheric and large but non-maximal solar mixing and a
small mixing angle $\theta_{13}$.
$SO(10)$ models have not yet been successful in explaining such
features, but one may try to meet such
challenges by considering the
possibility of an underlying flavor symmetry group $G$.
In GUT models based on the product group $SO(10) \times G$,
the three known fermion generations can be assigned to
representations of the flavor group $G$. Our choice of
$G$ is guided by the following considerations.
Since all irreps of abelian groups are one-dimensional,
only non-abelian groups are suited to explain the existence
of more than one fermion generation.
Furthermore, unlike continuous symmetries, the break-down of
discrete global symmetries does, in general, not give rise
to undesired Goldstone bosons. This suggests to stick
to non-abelian, discrete flavor groups.
Only a few $SO(10) \times G$ models, with $G$ being non-abelian
and discrete, have been studied so far.
For instance, models with $SO(10) \times S_4$~\cite{S4xSO10}
and $SO(10) \times A_4$~\cite{morisi}
symmetries have already been investigated.
In particular, models employing an $A_4$~\cite{A4,altarelli}
flavor symmetry may give tri-bimaximal leptonic mixing~\cite{HPS},
but, in general, in such models right- and left-handed fermion fields
transform differently under $A_4$. However, in $SO(10)$ GUTs
right- and left-handed fermion fields have to transform in the same
way under $A_4$~\cite{suitable},
since there all chiral matter fields of one generation
belong to the same $SO(10)$ irrep.
In~\cite{morisi} a non-supersymmetric
$SO(10) \times A_4$ model with type~I seesaw dominance has been
analyzed, which successfully preserves
tri-bimaximal leptonic mixing and can accommodate
all known fermion masses. The quark mixing angles, however,
are assumed to be zero.

In this paper, we investigate the fermionic sector of renormalizable
$SO(10) \times A_4$ GUT scenarios, with the three fermion families in
the three-dimensional irrep of $A_4$, while for the $SO(10)$ scalar
irreps occurring in Yukawa couplings we allow
all possible $A_4$ irreps. We do not discuss the difficult problem of
vacuum alignment, but rather assume that we can dispose of the VEVs
according to our needs. With this assumption, we will see that the
$SO(10) \times A_4$ structure enforces the building blocks of the fermion
mass matrices to consist of diagonal and off-diagonal matrices. A crucial
role will play the mass matrices of the down quarks and the charged
leptons. Requiring solely that the scenario is able to reproduce
down-quark masses and charged-lepton masses, singles out a unique case
with respect to the transformation of the scalars under
$SO(10) \times A_4$. In that unique viable scenario, under the
assumption of type~II seesaw dominance, we will find
the Zee--Wolfenstein form~\cite{zee-model,wolfenstein}
of the mass matrix of light neutrinos.
The well-known phenomenological problems~\cite{zee-pred} of
this mass matrix turn out to be completely resolvable by
contributions to lepton mixing from the charged-lepton sector.
We want to stress, however, that our usage of the $A_4$ flavor symmetry
does \emph{not} enforce tri-bimaximal mixing in the
lepton sector.

Though we have in mind a supersymmetrized scenario, supersymmetry
enters our considerations only via the fermion masses. For the
numerics we use masses at the GUT scale, which have been obtained
through the renormalization group equations of the MSSM.

Our paper is organized as follows.
In Section~\ref{sec-zee} we summarize
the properties of the Zee--Wolfenstein neutrino mass matrix.
The $SO(10)\times A_4$ GUT scenario is developed in
Section~\ref{sec-model}.
The methods and results of our numerical analysis
are discussed in Section~\ref{sec-numerics}.
Section~\ref{concl} is devoted to the conclusions.

\section{The Zee--Wolfenstein mass matrix in a nutshell}
\label{sec-zee}

The Zee model generates Majorana neutrino masses at the one-loop
level~\cite{zee-model,kummer}. Its neutrino mass matrix has,
in general, non-zero elements on the diagonal.
However, with a suitable $\mathbbm{Z}_2$ symmetry one can enforce
a vanishing diagonal in $\mnu$ at the one-loop level~\cite{wolfenstein}.
This Zee--Wolfenstein neutrino mass matrix
is a symmetric $3 \times 3$ matrix whose diagonal elements are zero:
\begin{equation}
\mnu = \left( \begin{array}{ccc}
0 & a & b \\
a & 0 & c \\
b & c & 0
\end{array} \right).
\label{Zee}
\end{equation}
From now on we will discuss only this case.
The restricted Zee model
has the property that one can make a basis transformation such that the
charged-lepton mass matrix is diagonal but the form of $\mnu$ given
by~(\ref{Zee}) persists. Thus, without loss
of generality, we will assume in this section that
the charged-lepton mass matrix is diagonal.

In diagonalizing the matrix~(\ref{Zee}), one can first remove the phases of
$\mnu$. These phases can be absorbed into
the charged-lepton fields. Thus we take
the matrix entries $a,b,c$ to be real.
Since the Zee--Wolfenstein
mass matrix is traceless and symmetric one has~\cite{he-zee}
\begin{equation}
\lambda_1 + \lambda_2 + \lambda_3 = 0,
\label{traceless}
\end{equation}
where the $\lambda_i$ denote the real eigenvalues of $\mnu$.
Writing $m_i$ for the masses of the light neutrinos,
we have $m_i = |\lambda_i|$.

In the case of inverted ordering
$m_3 < m_1 < m_2$ of the neutrino masses
($\deltasol = m_2^2 - m_1^2$, $\deltaatm = m_2^2 - m_3^2$),
it has been pointed out in Ref.~\cite{zee-pred}
that the mass matrix~(\ref{Zee})
together with $\deltasol \ll \deltaatm$ leads, for all practical purposes,
to maximal solar mixing $\theta_{12}=\pi/4$ and $\theta_{13}=0$.
Furthermore, the
atmospheric mixing angle $\theta_{23}$ can be \emph{chosen} to be maximal.
While the latter two properties are most welcome, maximal solar
mixing is excluded by more than $5\sigma$ by experimental data~\cite{schwetz1}.
In~\cite{balaji} it is has been shown that deviations
from maximal solar mixing are severely constrained through
\begin{equation}
|\cos 2\theta_{12}| \lesssim \frac{1}{4} \,
\frac{\deltasol}{\deltaatm}\,.
\end{equation}
The neutrino masses are approximately given by
\begin{equation}
\label{masses-inverted}
m_3 \simeq \frac{1}{2} \,
\frac{\deltasol}{\sqrt{\deltaatm}}
\,, \quad
m_1 \simeq m_2 \simeq \sqrt{\deltaatm}\,.
\end{equation}
Thus one obtains $m_3 \ll m_1 \simeq m_2$ and the resulting neutrino mass
spectrum exhibits an inverted hierarchy.

In the case of the normal ordering
$m_1 < m_2 < m_3$ of the neutrino masses
($\deltasol = m_2^2 - m_1^2$, $\deltaatm = m_3^2 - m_1^2$)
things are even worse. Although it is now possible to have maximal
atmospheric mixing and, at the same time, allowing the solar mixing
angle to be in perfect agreement with experimental data, the mixing
angle $\theta_{13}$ turns out to be much too large~\cite{zee-pred}:
\begin{equation}
\sin^2\theta_{13} \simeq \frac{1}{3}\,.
\end{equation}
The neutrino mass spectrum can be estimated by
\begin{equation}
\label{masses-normal}
m_1 \simeq m_2 \simeq \frac{1}{2}\,m_3 \simeq%
\sqrt{\frac{\deltaatm}{3}}\,.
\end{equation}
Therefore, all neutrino masses will be of the same
order of magnitude, but the mass spectrum cannot be quasi-degenerate.

Concerning neutrinoless double-$\beta$ decay,
the relevant observable is the effective Majorana neutrino mass
$\meff \equiv \left|\sum_i U^2_{ei}m_i \right|$, where $U$ denotes the
unitary leptonic mixing (PMNS) matrix. The mass
$\meff$ is equal to the modulus
of the $\left( e,e \right)$ matrix element of $\mnu$,
which is exactly zero in the Zee--Wolfenstein case. Thus the model prevents
neutrinoless double-$\beta$ decay.

In summary, the Zee--Wolfenstein model is not viable because it does not
give a consistent explanation of \emph{all} current experimental data of
the neutrino sector (\emph{i.e.}\ two mass-squared differences plus
three mixings angles). It is the purpose of this paper to embed the
Zee--Wolfenstein neutrino mass matrix in an $SO(10)$ GUT. In such an
environment, the zeros in the diagonal of $\mnu$ are \emph{not} stable under a
basis change such that the charged-lepton mass matrix becomes
diagonal. Therefore, as we will show, contributions
from the charged-lepton sector can provide the
necessary remedy for correcting the too large mixing angle
$\theta_{12}$ in the case of inverted hierarchy and $\theta_{13}$ in the case
of normal hierarchy~\cite{bimaximal-deviation}.
As an additional bonus, a non-vanishing $\meff$
and, therefore, neutrinoless double-$\beta$ decay becomes possible.

\section{The $SO(10)\times A_4$ model}
\label{sec-model}

The tensor product of the $SO(10)$ spinor representation of the fermions is
given by~\cite{sakita,slansky}
\begin{equation}
\label{SO10tensor}
\mathbf{16} \otimes \mathbf{16}
= \left( \mathbf{10} \oplus \mathbf{126} \right)_\mathrm{S}
\oplus \mathbf{120}_\mathrm{AS},
\end{equation}
where the subscripts S and AS refer to symmetric and antisymmetric Yukawa
coupling matrices, respectively. Renormalizable $SO(10)$ GUTs can generate
fermion masses at the tree level only by the scalar irreps
$\ten$, $\oht$ and $\hts$.

The 12-element group $A_4$ is popular as a family symmetry in model
building--- see~\cite{A4} for a selection of the vast $A_4$ literature and
\cite{altarelli} for a review on the group $A_4$ and models.
It has three one-dimensional
irreps and one three-dimensional irrep. The tensor product
$\mathbf{3} \otimes \mathbf{3}$ contains all
one-dimensional irreps exactly once, but the
three-dimensional irrep is contained
twice. While the Yukawa couplings corresponding to the one-dimensional irreps
are diagonal and, therefore, symmetric, the couplings of the
$\mathbf{3} \oplus \mathbf{3} \in \mathbf{3} \otimes \mathbf{3}$
are off-diagonal, but no special symmetry
property is fixed. However, Eq.~(\ref{SO10tensor}) suggests to choose one
three-dimensional irrep with symmetric and the other one with
antisymmetric tensor indices:
\begin{equation}
\label{A4tensor}
\mathbf{3} \otimes \mathbf{3}
= \left( \ones \oplus \mathbf{3} \right)_\mathrm{S} \oplus
\mathbf{3}_\mathrm{AS}.
\end{equation}

Now we consider $SO(10) \times A_4$ and investigate possible Yukawa couplings
and fermion mass matrices under the assumption that the fermions transform as
$\mathbf{16} \otimes \mathbf{3}$, which is is clearly the only
reasonable choice if
we want to take advantage of the non-abelian character of $A_4$.
Equations~(\ref{SO10tensor}) and (\ref{A4tensor}) dictate that the $\oht$ can
only transform as a $\mathbf{3}$ under $A_4$, while for the for $\ten$ and
$\hts$ singlet \emph{and} triplet irreps of $A_4$ are possible.
Let us consider the case where the scalars responsible for Yukawa couplings
transform as
\begin{equation}\label{case}
\ten \otimes (\ones) \quad \mbox{and} \quad \hts \otimes \mathbf{3}.
\end{equation}
Then, in a symbolic way, writing down only the $A_4$ part,
the Yukawa couplings are given by
\begin{align}
\label{Y10}
& \sum_{i=1}^3 h_i \sum_{a=1}^3 \omega^{(i-1)(a-1)} \mathbf{16}_a
\mathbf{16}_a \ten_i \,+
\\ \label{Y126bar} &
(\st_2 \st_3 + \st_3 \st_2)\, \hts_1 +
(\st_3 \st_1 + \st_1 \st_3)\, \hts_2 +
(\st_1 \st_2 + \st_2 \st_1)\, \hts_3,
\end{align}
where $a$ is a family index and $\omega = (-1 + i \sqrt{3})/2$.
Furthermore, we make two assumptions:
\begin{enumerate}
\renewcommand{\labelenumi}{\roman{enumi})}
\item
All VEVs which occur in the
scalars can have independent values.
\item
Type II seesaw dominates in the neutrino mass matrix.
\end{enumerate}
These assumptions together with Eq.~(\ref{case}) define the scenario
we will investigate in the following.

We furthermore assume that our models can be extended in a suitable
way to solve the doublet-triplet splitting 
problem.\footnote{For instance, the
Dimopoulous--Wilczek mechanism~\cite{doublet-triplet1}
and the missing partner mechanism~\cite{doublet-triplet2}
provide viable solutions of the doublet-triplet
splitting problem in $SO(10)$ GUTs.}
Moreover, since we have in mind the MSSM, with only two Higgs doublets, 
as the low-energy limit of our $SO(10)$ models we must assume a
suitable doublet-doublet splitting as well, which is usually achieved
by finetuning~\cite{minimalfinetuning}.
These assumptions are not innate to the models presented here but are
well-known problems in GUTs.

Let us now derive some consequences of our scenario.
Because of assumption i), the Yukawa
couplings~(\ref{Y10}) produce \emph{diagonal} mass terms with three independent
entries, one for the up-quark mass matrix ($q=u$) and another one for
the down quark mass matrix ($q=d$):
\begin{equation}
\mbox{diag} \left(
h_1 v_{q1} + h_2 v_{q2} + h_3 v_{q3}, \,
h_1 v_{q1} + \omega h_2 v_{q2} + \omega^2 h_3 v_{q3}, \,
h_1 v_{q1} + \omega^2 h_2 v_{q2} + \omega h_3 v_{q3} \right),
\end{equation}
where the $v_{qi}$ are the VEVs appearing in the scalar 10-plets.
Next we consider the Yukawa couplings~(\ref{Y126bar}).
Again because of assumption i), this Yukawa
interaction generates two independent \emph{off-diagonal} contributions to
the mass matrices of up and down quarks.

Studying the system of mass matrices, we find that
assumptions i) and ii) lead to a decoupling of the up-quark
mass matrix $M_u$ from the rest of the system.
This is so because the quark mass matrices are given by
$M_u = H' + F'$ and $M_d = H + F$,
where $H'$ and $H$ are independent diagonal matrices, while $F'$ and
$F$ are independent off-diagonal matrices. Therefore, $M_u$ would
only be related to the system of mass matrices through
the neutrino Dirac-mass matrix $M_D = H' - 3\, F'$, but
this relationship is irrelevant due to assumption ii).
Since $M_u$ is a general symmetric matrix independent of
the rest of the system of mass matrices, the CKM matrix can always be
reproduced. The other side of the coin is that our scenario loses
predictivity because it is neither
restricted by the values of the up-quark masses nor
by the experimental information on the CKM matrix.

The remaining system of mass matrices which we want to study consists
of the mass matrices of down-type quarks and
charged-leptons, given by
\begin{equation}
\label{Md-Ml}
M_d = H + F \quad \mbox{and} \quad M_\ell = H - 3\,F,
\end{equation}
respectively, where $H$ is diagonal,
while $F$ is off-diagonal, and of the neutrino mass
matrix $\mnu$ of Eq.~(\ref{Zee}). Without loss of generality,
$H$ can be assumed to be real, but $F$ and $\mnu$ have complex
entries.\footnote{Now we cannot absorb the phases of $\mnu$ into the
  charged-lepton fields since $M_\ell$ is not diagonal.}
Note that in view of assumption i)
the entries in $\mnu$ are independent of $F$, but $M_\ell$ and
$\mnu$ are coupled via the PMNS matrix
\begin{equation}\label{PMNS}
U = U_\ell^\dagger U_\nu
\quad \mbox{with} \quad
U_\ell^T M_\ell U_\ell = \mbox{diag}\, (m_e,m_\mu,m_\tau),
\quad
U_\nu^T \mnu U_\nu = \mbox{diag}\, (m_1,m_2,m_3).
\end{equation}
Counting the number of parameters, we find nine absolute values
and five phases,\footnote{Of the three phases in $\mnu$ one can be
  removed.} while the number of observables to be fitted is 11:
three charged-lepton masses, three down-quark masses, two neutrino
mass-squared differences and three lepton mixing angles.
The fitting procedure and predictions of our scenario will be exposed
in the next section.

We note that the family symmetry $A_4$ has the effect of generating
independent diagonal and off-diagonal contributions to the quark and
lepton mass matrices.
Adhering to the two assumptions presented above but using
other $A_4$ representations than those of Eq.~(\ref{case}), we can
find several other scenarios.
E.g., with 120-plets, antisymmetric
off-diagonal mass matrix contributions are generated. A
list of such cases is presented in Table~\ref{models}. There we
confine ourselves to a maximum of three scalars per $SO(10)$ irrep,
the 126-plet must always be present to allow a viable type~II seesaw
neutrino mass matrix and the $\ten$ and $\oht$ are not present at the
same time; the latter condition is for avoiding a proliferation of
parameters. However, it will turn out that the only viable scenario is
the one defined via Eq.~(\ref{case}).

\section{The numerical analysis}
\label{sec-numerics}
\begin{table}[t]
\begin{center}
\renewcommand{\arraystretch}{1.2}
\begin{tabular}{cc}
\begin{tabular}[t]{|c|c|} \hline
\multicolumn{2}{|c|}{Leptons} \\ \hline\hline
$m_e$       &
$0.3585 \berr 0.0003 \tr 0.0003 \eerr$ \\ \hline
$m_\mu$     &
$75.6715 \berr 0.0578 \tr 0.0501 \eerr$ \\ \hline
$m_\tau$    &
$1292.2 \berr 1.3 \tr 1.2 \eerr$ \\ \hline
$\deltasol$ & $(7.9 \pm 0.3) \times 10^{-5}$ \\ \hline
$\deltaatm$ & $\Big(2.50 \berr 0.20 \tr 0.25 \eerr
\Big) \times 10^{-3}$ \\ \hline
$s_{12}^2$  & $0.31 \pm 0.025$ \\ \hline
$s_{23}^2$  & $0.50 \pm 0.065$ \\ \hline
$s_{13}^2$  & $< 0.0155$ \\ \hline
\end{tabular}
&
\begin{tabular}[t]{|c|c|} \hline
\multicolumn{2}{|c|}{Quarks} \\ \hline\hline
$m_d$ &
$1.03 \pm 0.41$ \\ \hline
$m_s$ & $19.6 \pm 5.2$ \\ \hline
$m_b$ & $1063.6 \berr 141.4 \tr 086.5 \eerr$ \\ \hline
\end{tabular}
\end{tabular}
\end{center}
\caption{Input data (central values and $1\sigma$ errors)
at the GUT scale of $M_\mathrm{GUT} = 2 \times
10^{16} \: \mbox{GeV}$ for $\tan\beta = 10$. The charged-fermion masses are
taken from~\cite{das}, except for the values of $m_d$ and $m_s$;
these were obtained by taking their low-energy values from~\cite{RPP} and
scaling them to $M_\mathrm{GUT}$. As for $\deltaatm$, we use the value obtained
in~\cite{schwetz1}. We have copied the remaining input from Table~I
in~\cite{schwetz}. Charged-fermion masses are in units of MeV,
neutrino mass-squared differences in $\mbox{eV}^2$. We have used the
abbreviations $s^2_{12} \equiv \sin^2 \theta_{12}$, \emph{etc}.
The angles in the left table refer to the PMNS matrix.\label{input}}
\end{table}
We perform a global $\chi^2$~analysis of the $SO(10)\times A_4$
scenario defined by Eq.~(\ref{case}) and assumptions i) and ii)
by employing the downhill simplex method~\cite{downhill}.
In Table~\ref{input}
the observable quantities $O_i$ are specified in the form
\begin{equation}
O_i=\bar{O}_i\pm\sigma_i,
\end{equation}
where $\bar{O}_i$ and $\sigma_i$ denote central values and
$1\sigma$ deviations, respectively. The index $i=1,\ldots,11$ labels
the different observables given in Table~\ref{input}.
The masses in that table refer to the mass values at a GUT scale
of $2 \times 10^{16}$ GeV, obtained via the renormalization group
equations of the MSSM, for the ratio of Higgs doublet VEVs $\tan \beta
= 10$.\footnote{Since we do not have quasi-degenerate
  neutrino mass spectra, the effect of the renormalization group
  running on the lepton mixing angles is negligible~\cite{antusch}.}
Writing $\mathbf{x}$ for the set the 14 model parameters and
$P_{i}(\mathbf{x})$ for the resulting model predictions, one can
define a $\chi^{2}$ function by
\begin{equation}
\label{chisquare}
\chi^{2}(\mathbf{x})=\sum_{i=1}^{11}\left(\frac{P_i(\mathbf{x})-\bar{O}_i}%
{\sigma_i}\right)^{2}.
\end{equation}
The global minimum of $\chi^{2}$ will represent the best possible agreement
of theoretical predictions and experimental data.
This minimization task is performed using the
downhill simplex method.

For investigating the variation of $\chi^2$ as a function of the value
$\widehat O$ of an observable $O$, we add the ``pinning term''
$(P(\mathbf{x})-\widehat{O})^2/(0.01\,\widehat{O})^2$ to $\chi^2$,
where $P(\mathbf{x})$ represents the theoretical prediction for $O$.
Note that if $O$ agrees with one of observables $O_i$ occurring
in $\chi^2$ of Eq.~(\ref{chisquare}), the $O_i$ term has to be
removed from Eq.~(\ref{chisquare}).
The small error in the denominator of the ``pinning term''
guarantees to pin the observable $O$ down to the value $\widehat{O}$.
The pinning procedure performs as
desired when the contribution of the pinning term to $\chi^{2}$ is
negligible.

\begin{table}[t]
\begin{center}
\renewcommand{\arraystretch}{1.2}
\begin{tabular}{|c|c|c|c|c|}
\hline
Model & $\ten$ & $\oht$ & $\hts$ & $\chi_{d\ell}^{2}$ \\ \hline\hline
A & $\mathbf{3}$ & $-$ & $\mathbf{3}$ & $10^{6}$ \\ \hline
B & $-$ & $\mathbf{3}$ & $\mathbf{3}$ & $46$ \\ \hline
C & $\mathbf{3}$ & $-$ & $\ones$ & $46$ \\ \hline
D & $-$ & $\mathbf{3}$ & $\ones$ & $46$ \\ \hline
E' & $\mathbf{1}$ & $-$ & $\mathbf{3}$ & $7\times 10^{4}$ \\ \hline
E & $\ones$ & $-$ & $\mathbf{3}$ & $0.89$ \\ \hline
\end{tabular}
\end{center}
\caption{A variety of renormalizable GUT models based on $SO(10)\times A_4$.
Each line corresponds to a distinctive model scenario.
Columns 2--4 specify the transformation properties of the $SO(10)$
scalar multiplets
under the flavor symmetry group $A_4$. The last column gives the
best-fit values $\chi_{d\ell}^{2}$ when fitting charged-lepton
and down-type quark masses.
\label{models}}
\end{table}
As mentioned at the end of Section~\ref{sec-model}, we have not only
investigated the scenario defined by Eq.~(\ref{case}) but also a
variety of others which are characterized by the $A_4$ transformation
properties of their scalar $SO(10)$ multiplets in columns 2--4 of
Table~\ref{models} (models A--E'). We have found that all these
scenarios fail already to reproduce the down-quark and charged-lepton
masses---see the value of the corresponding $\chi^2_{d\ell}$ in the
last column of Table~\ref{models}.\footnote{For case A this failure is
trivial: $M_d$ and $M_\ell$ are symmetric with a vanishing diagonal,
therefore, Eq.~(\ref{traceless}) holds, which is in contradiction to
the strong hierarchy in the down-quark and charged-lepton masses.}
For comparison we have also
presented the $\chi^2_{d\ell}$ of our successful scenario in the line
labeled by E, which will be investigated in the rest of this paper.

\subsection{Predictions for the case of normal neutrino mass ordering}
\label{sec-normal}
We search for the best-fit solution for the normal neutrino mass spectrum
$m_1 < m_2 < m_3$. In this case we find an excellent fit with the
following properties:
\begin{equation}
\label{bf-normal-results}
\begin{array}{l}
\chi^2 = 0.96, \\[1mm]
m_1 = 2.795 \times 10^{-2} \: \mbox{eV}, \quad
m_2 = 2.933 \times 10^{-2} \: \mbox{eV}, \quad
m_3 = 5.728 \times 10^{-2} \: \mbox{eV}.
\end{array}
\end{equation}
The corresponding values of the matrix elements
of $H$, $F$, $\mnu$ are given by
\begin{eqnarray}
\label{bf-normal-params}
H & = & \left( \begin{array}{ccc}
14.8905 & 0 & 0 \\
0 & 14.9798 & 0 \\
0 & 0 & 1189.49
\end{array} \right), \nonumber \\
F & = & \left( \begin{array}{ccc}
0 & 4.45699\,e^{i\,0.990889\pi} & 83.3159\,e^{-i\,0.940907\pi} \\
4.45699\,e^{i\,0.990889\pi} & 0 & 86.5511\,e^{i\,0.946557\pi} \\
83.3159\,e^{-i\,0.940907\pi} & 86.5511\,e^{i\,0.946557\pi} & 0
\end{array} \right), \\
\mnu & = & \left( \begin{array}{ccc}
0 & 2.83435\,e^{i\,0.267950\pi} & 2.93292\,e^{i\,0.649567\pi} \\
2.83435\,e^{i\,0.267950\pi} & 0 & 2.82486\,e^{i\,0.5\pi} \\
2.93292\,e^{i\,0.649567\pi} & 2.82486\,e^{i\,0.5\pi} & 0
\end{array} \right)\times 10^{-2}, \nonumber
\end{eqnarray}
where the numerical values in $H$ and $F$ are in units of
MeV, while the entries in $\mnu$ are in units of eV.

The non-zero value of $\chi^2$ stems from the
deviation of the bottom-quark mass $m_b$ from its
central value by $+0.98\sigma$.
The remaining observables of Table~\ref{input} are
fitted perfectly. Thus the model succeeds in correcting
the too large value for the mixing angle $\theta_{13}$ of
the Zee--Wolfenstein model, despite
the close relationship between $M_d$ and $M_\ell$ given by
Eq.~(\ref{Md-Ml}) which, on the other hand,
leads to the desired unification of
$m_b$ and $m_{\tau}$, as
will be discussed in Section~\ref{sec-unification}.

As explained in Section~\ref{sec-zee},
the three light neutrino
masses cannot be independent of each other. The sum
of the eigenvalues of $\mnu$ must be zero,
which translates into $m_1 + m_2 - m_3 = 0$.
This can easily be verified for the
neutrino masses of the best-fit~(\ref{bf-normal-results}).
The sum of the neutrino masses is
$\Sigma \equiv \sum_i m_i = 2\,m_3 = 0.11 \: \mbox{eV}$,
which lies safely below the cosmological bound
$\Sigma \lesssim 1 \: \mbox{eV}$~\cite{raffelt}.

The neutrino mass spectrum has to fulfill the approximate
relation~(\ref{masses-normal}). Inserting the
central value for $\deltaatm$ from Table~\ref{input} into
Eq.~(\ref{masses-normal}) gives
$m_1 \simeq m_2 \simeq 2.89 \times 10^{-2} \: \mbox{eV}$ and
$m_3 \simeq 5.77 \times 10^{-2} \: \mbox{eV}$, which is in
good agreement with the above best-fit results.

The quantity $R \equiv m_1/\sqrt{\deltasol}$ measures how hierarchical
a neutrino mass spectrum is. $\chi^2$ as a function of $R$ is
depicted in the right panel of Figure~\ref{fig_neutrino}.
We read off that $R \sim 3.1$ is preferred and for the values
$2.4 \lesssim R \lesssim 3.7$ one obtains fits with $\chi^2\lesssim15$.
Thus the mass spectrum is neither hierarchical nor
quasi-degenerate,\footnote{Typically, quasi-degenerate neutrino spectra
would correspond to $R \gtrsim \sqrt{\deltaatm/\deltasol} \simeq 5.6$\,.}
but is located between these extrema. The narrow range of allowed values
for $R$ reflects the clear-cut prediction of the Zee--Wolfenstein mass
matrix for the neutrino mass spectrum.

Figure~\ref{fig_neutrino}~(left panel) shows the constraints on the
atmospheric mixing angle $\theta_{23}$. One can see that
values of $\sin^2\theta_{23}$ smaller than 0.38 ($-2\sigma$) are strongly
disfavored and thus a strict lower bound for $\theta_{23}$
is established. However, very good fits are also possible
for values of $\sin^2\theta_{23}$ significantly larger than
the best-fit value of 0.5.

Concerning the solar mixing angle, Figure~\ref{fig_neutrino}~(middle panel)
shows that the whole physically allowed range for $\sin^2\theta_{12}$
gives excellent fits and therefore no prediction can be obtained.

Regarding the mixing angle $\theta_{13}$, the best-fit solution gives a value
of $\sin^2\theta_{13}=2\times10^{-4}$. However, also significantly smaller
(down to $10^{-6}$) and larger values (up to 0.1) for $\sin^2\theta_{13}$
are equally allowed. Thus the severe problem of the original
Zee--Wolfenstein mass matrix $(\sin^2\theta_{13} \simeq 1/3)$
can be resolved completely by contributions from the charged-lepton sector.

The best-fit gives $\delta_\mathrm{PMNS} = 31^\circ$ for the leptonic
CP phase. However, varying $\delta_\mathrm{PMNS}$ shows that the
whole $[0^\circ,360^\circ]$ range allows for very good fits and therefore
no prediction can be made.

The effective Majorana mass of neutrinoless
double-$\beta$ decay $\meff$ for the normal spectrum is given by
\begin{eqnarray}
\label{m2beta-normal}
\meff & = & \left| \left( m_1 \, c_{12}^2 +
\sqrt{m_1^2 + \deltasol}\, s_{12}^2\, e^{i\beta_1} \right) c_{13}^2 +
\right. \nonumber \\[1mm] && \left.
\sqrt{m_1^2 + \deltaatm}\, s_{13}^2\, e^{i\beta_2} \right|,
\end{eqnarray}
where $\beta_1$ and $\beta_2$ are Majorana phases.
Here and in the following we use the abbreviations
$c_{12} \equiv \cos \theta_{12}$, $s_{12} \equiv \sin \theta_{12}$,
\emph{etc}.
After inserting into Eq.~(\ref{m2beta-normal}) $m_1$ from the
best-fit~(\ref{bf-normal-results}),
employing for the other parameters the corresponding
central values from Table~\ref{input},
and varying the two phases $\beta_1$ and $\beta_2$ \emph{freely} between
$0^\circ$ and $360^\circ$,
we obtain the bounds
\begin{equation}
\label{meff-bounds-normal}
10.2 \: \mbox{meV}
\leq \meff \leq
28.4 \: \mbox{meV}.
\end{equation}
%
On the other hand, the phases $\beta_1$ and $\beta_2$ are actually
functions of the parameters of our scenario and are determined by the
fit. Using the best-fit parameters~(\ref{bf-normal-params}) for the
calculation of the effective Majorana mass,
we obtain $\meff = 28.4 \: \mbox{meV}$, which
is identical to the upper bound in~(\ref{meff-bounds-normal}).
Figure~\ref{fig_2beta} presents the change of $\chi^2$ when
$\meff$ is varied.
We can read off that the range for the effective mass
is much more restricted than~(\ref{meff-bounds-normal})
would suggest. Obviously, the increase of $\chi^2$ for larger
values of $\meff$ is caused by exceeding the
upper bound of~(\ref{meff-bounds-normal}).
The strong increase of $\chi^2$ for smaller values of $\meff$,
however, is a clear-cut model prediction.
For instance, allowing for only moderate good fits
with $\chi^2 \lesssim 5$ results in the severely restricted range
$25 \: \mbox{meV}
\lesssim \meff \lesssim
31 \: \mbox{meV}$,
which could be tested by future neutrinoless double-$\beta$ decay
experiments sensitive to $\meff \gtrsim 10 \: \mbox{meV}$.

\subsection{Predictions for the case of inverted neutrino mass ordering}
\label{sec-inverted}
\begin{figure}[t]
\begin{center}
\epsfig{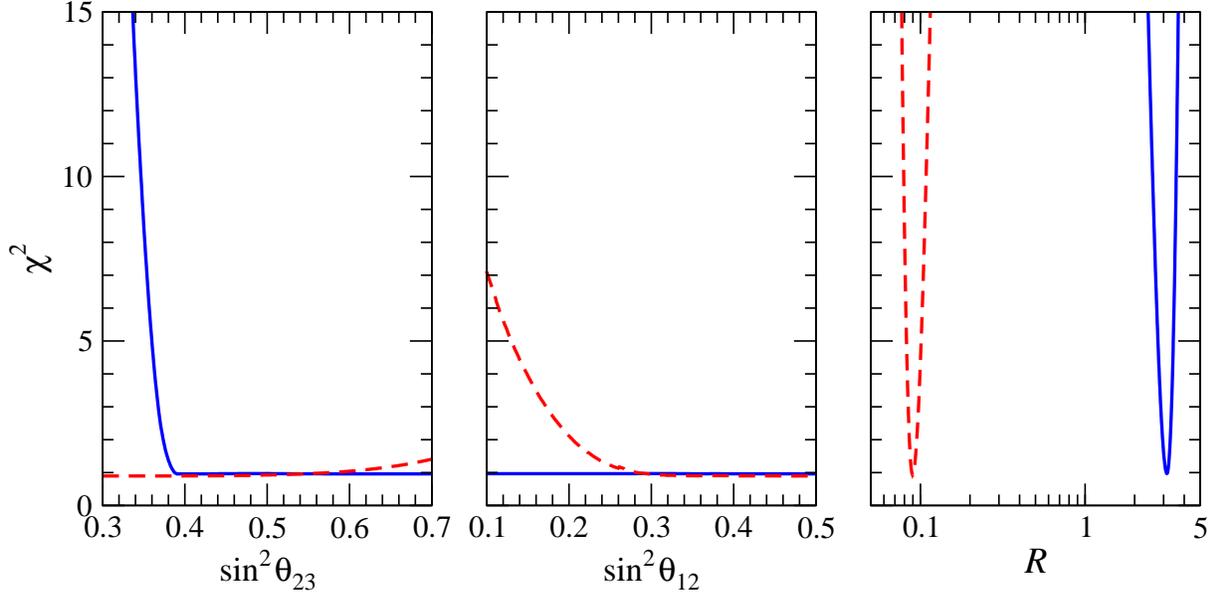}
\end{center}
\caption{$\chi^2$ as a function of $\sin^2\theta_{23}$ (left panel),
$\sin^2\theta_{12}$ (middle panel) and of
$R \equiv m_\mathrm{min}/\sqrt{\deltasol}$ (right panel),
where $m_\mathrm{min} = m_1$ for the normal and $m_3$ for the inverted
neutrino mass spectrum.
The solid lines correspond to normal neutrino mass ordering,
while the dashed lines refer to the inverted neutrino mass spectrum.
\label{fig_neutrino}}
\end{figure}

\begin{figure}[t]
\begin{center}
\epsfig{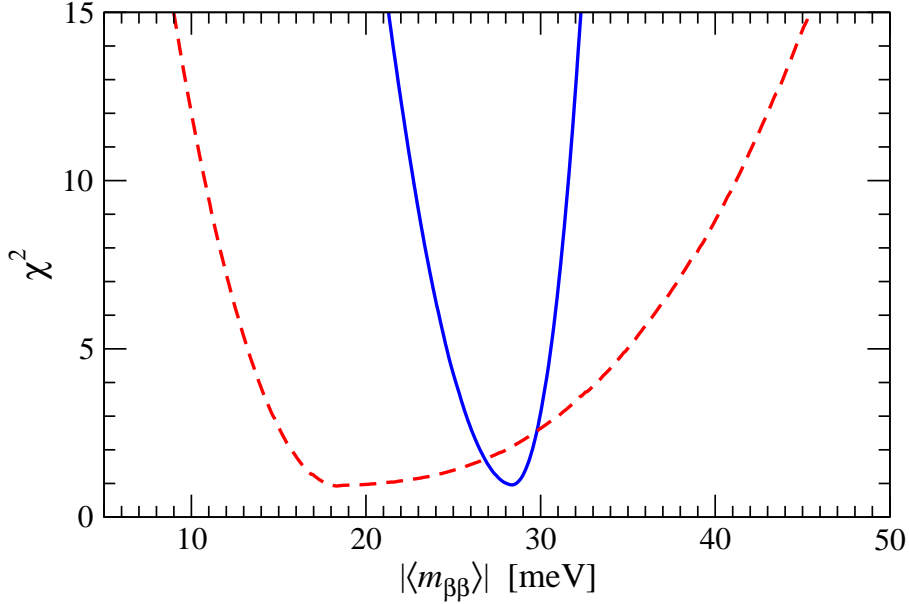}
\end{center}
\caption{$\chi^2$ as a function of the effective Majorana mass
$\meff$ probed in neutrinoless double-$\beta$ decay experiments.
The solid line corresponds to normal neutrino mass ordering,
while the dashed line refers to the inverted neutrino mass spectrum.
\label{fig_2beta}}
\end{figure}

The best-fit solution for the inverted neutrino mass spectrum
$m_3 < m_1 < m_2$ turns out to be excellent as well.
It is characterized by the following properties:
\begin{equation}
\label{bf-inverted-result}
\begin{array}{l}
\chi^2 = 0.92, \\[1mm]
m_1 = 4.921 \times 10^{-2} \: \mbox{eV}, \quad
m_2 = 5.000 \times 10^{-2} \: \mbox{eV}, \quad
m_3 = 7.963 \times 10^{-4} \: \mbox{eV},
\end{array}
\end{equation}
with the matrices
\begin{eqnarray}
\label{bf-inverted-params}
H & = & \left( \begin{array}{ccc}
8.24821 & 0 & 0 \\
0 & 23.3969 & 0 \\
0 & 0 & 1184.52
\end{array} \right), \nonumber \\
F & = & \left( \begin{array}{ccc}
0 & 6.34572\,e^{i\pi} & 39.2455 \\
6.34572\,e^{i\pi} & 0 & 116.271 \\
39.2455 & 116.271 & 0
\end{array} \right), \\
\mnu & = & \left( \begin{array}{ccc}
0 & 4.32801 & 2.42291\,e^{i\pi} \\
4.32801 & 0 & 0.0934209\,e^{i\pi} \\
2.42291\,e^{i\pi} & 0.0934209\,e^{i\pi} & 0
\end{array} \right)\times 10^{-2}, \nonumber
\end{eqnarray}
where the numerical values in $H$ and $F$ are in units of
MeV, while the entries in $\mnu$ are in units of eV.

The $\chi^2$~analysis reveals that the removal of the
non-trivial complex phases from $F$ and $\mnu$ does not affect the
goodness of the fit. Thus we specified here the fitting parameters for the
CP conserving case.\footnote{For the normal neutrino spectrum, however,
the CP conserving case results in a worse, but still very good fit
with $\chi^2=1.94$. Here, the main contributions to $\chi^2$ stem
from $m_b$ ($+1.23\sigma$) and $\sin^2\theta_{23}$ ($-0.64\sigma$)\,.}
However, the subsequent numerical analysis is performed with
the inclusion of the five phase parameters (CP non-conservation).
As in the case of normal neutrino mass ordering,
the main contribution to $\chi^2$ is caused by the
bottom-quark mass $m_b$, being too large by $0.95\sigma$.
All the other observables are fitted very accurately.
Thus the GUT model allows for a considerably reduction of
the maximal solar mixing angle $\theta_{12}$, which
spoiled the Zee--Wolfenstein model.

$\mnu$ being of the Zee--Wolfenstein form implies
$m_1 - m_2 + m_3 = 0$ for the three light neutrino masses.
Taking the neutrino masses from the best-fit~(\ref{bf-inverted-result}),
we get $\Sigma \equiv \sum_i m_i = 2\,m_2 =0.10 \: \mbox{eV}$,
which is safely below the cosmological limit~\cite{raffelt}.
Inserting the central values for the mass-squared differences
from Table~\ref{input} into Eqs.~(\ref{masses-inverted}) gives
$m_1 \simeq m_2 \simeq 5 \times 10^{-2} \: \mbox{eV}$ and
$m_3 \simeq 7.9 \times 10^{-4} \: \mbox{eV}$, which is in
good agreement with the numerically obtained
best-fit values~(\ref{bf-inverted-result}).

$\chi^2$ as a function of $R \equiv m_3/\sqrt{\deltasol}$ is shown in
Figure~\ref{fig_neutrino}~(right panel). We can read off that
$R \sim 0.09$ is preferred and for the values
$0.077 \lesssim R \lesssim 0.11$ one gets fits with $\chi^2\lesssim15$.
As in the case of normal neutrino mass ordering,
the range for $R$ is very restricted.
Hierarchy is strongly preferred, however, too small values for $m_3$ are
strictly forbidden.

Figure~\ref{fig_neutrino}~(middle panel) depicts the constraints on the
solar mixing angle $\theta_{12}$. We can read off that
values for $\sin^2\theta_{12}$ smaller than 0.3 become increasingly disfavored.
However, very good fits can also be found for values of
$\sin^2\theta_{12}$ larger than the best-fit value,
and maximal solar mixing also represents a very good fit.

Regarding the atmospheric mixing angle $\theta_{23}$,
Figure~\ref{fig_neutrino}~(left panel)
reveals that the whole physically allowed range for $\sin^2\theta_{23}$
gives very good fits and therefore no prediction can be obtained.
This property is seemingly a legacy of the original Zee--Wolfenstein model,
where the atmospheric mixing angle for inverted neutrino
mass ordering is unconstrained~\cite{zee-pred}.

As for the mixing angle $\theta_{13}$,
we find $\sin^2\theta_{13}=2.5\times10^{-3}$ for the best fit.
However, pinning $\sin^2\theta_{13}$ in $\chi^2$ shows that also smaller
(down to $10^{-6}$) and larger values (up to 0.1) are possible. For instance,
enforcing $\sin^2\theta_{13}=10^{-6}$ still allows $\chi^2=3.1$.

Concerning the leptonic CP phase $\delta_\mathrm{PMNS}$,
the specified best-fit, which employs only trivial complex phases,
gives $\delta_\mathrm{PMNS} = 180^\circ$.
However, varying $\delta_\mathrm{PMNS}$
in the range $(90^\circ,270^\circ)$ allows for fits of equally good
quality. Only
the neighborhood of $\delta_\mathrm{PMNS} \simeq  0^\circ$ seems to be
slightly disfavored by $\chi^2 \simeq 3.2$.

The effective Majorana mass for the
neutrinoless double-$\beta$~decay is given by
\begin{eqnarray}
\label{m2beta-inverted}
\meff & = & \left| \left( \sqrt{m_3^2 +
\deltaatm - \deltasol} \,
c_{12}^2\, e^{i\beta_1} + \right. \right.
\nonumber \\[1mm] && \left. \left.
\sqrt{m_3^2 + \deltaatm}\, s_{12}^2\, e^{i\beta_2}
\right) c_{13}^2 +
m_3\, s_{13}^2 \right|.
\end{eqnarray}
With $m_3$ from the best-fit and taking for the other parameters
in~(\ref{m2beta-inverted}) the corresponding central values in
Table~\ref{input}, free variation of the two
complex phases results in the following bounds on
the effective Majorana neutrino mass:
\begin{equation}
\label{meff-bounds-inverted}
18.5 \: \mbox{meV}
\leq \meff \leq
49.5 \: \mbox{meV}.
\end{equation}
On the other hand,
employing the best-fit parameters~(\ref{bf-inverted-params}),
we obtain $\meff = 18.4 \: \mbox{meV}$,
which is located close to
the lower bound of~(\ref{meff-bounds-inverted}).

Figure~\ref{fig_2beta} shows the change of $\chi^2$ under
variations of $\meff$.
We can see that the range for the effective mass
is less restricted than in the case of normal neutrino spectrum.
Clearly, the strong increase of $\chi^2$ for smaller
values of $\meff$ comes from falling below the lower bound
of~(\ref{meff-bounds-inverted}). The rise of $\chi^2$ is
less dramatic when moving to larger values of $\meff$.
However, there is a clear bias towards
values of $\meff$ in the lower half of the
range spanned by~(\ref{meff-bounds-inverted}).
We can also read off from Figure~\ref{fig_2beta} that
allowing moderately good fits with
$\chi^2 \lesssim 5$ gives
$13 \: \mbox{meV} \lesssim \meff \lesssim 35 \: \mbox{meV}$.
Moreover, we can see that
the $\meff$ regions where $\chi^2 \gtrsim 2$
are overlapping for both neutrino mass orderings
and thus one cannot discriminate with $\meff$ between
normal and inverted mass spectrum
in the overlap region.
However, $\meff \lesssim 20 \: \mbox{meV}$
(which is preferred) or
$\meff \gtrsim 33 \: \mbox{meV}$ is only possible for an inverted
hierarchy in our scenario.

\subsection{$b-\tau$ unification}
\label{sec-unification}
\begin{figure}[t]
\begin{center}
\epsfig{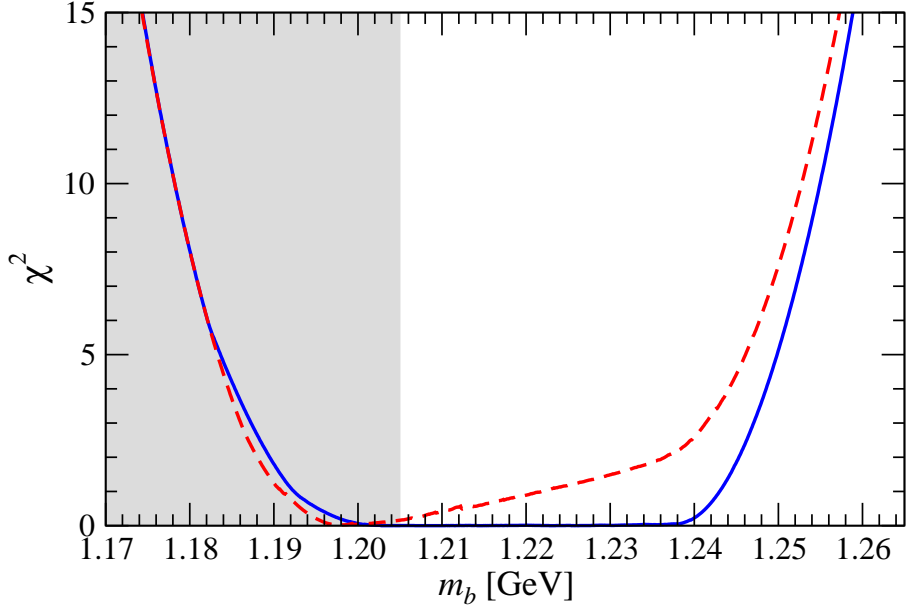}
\end{center}
\caption{$\chi^2$ as a function of the bottom-quark mass $m_b$.
The solid line corresponds to normal neutrino mass ordering,
while the dashed line refers to the inverted neutrino mass spectrum.
The shaded region indicates the $1\sigma$ interval for $m_b$
from Table~\ref{input}.
\label{fig_mb}}
\end{figure}

As has already been
noticed in Sections~\ref{sec-normal} and \ref{sec-inverted},
both best-fit values of $m_b$ are located near the upper $1\sigma$
bound of its input value from Table~\ref{input}.
The ratio $m_b/m_\tau$, using the mean values from
Table~\ref{input}, is $0.82$.
Employing the best-fit values for $m_b$ and $m_\tau$,
this ratio is higher, namely $m_b/m_\tau = 0.93$ for both normal
and inverted neutrino mass ordering.

Figure~\ref{fig_mb} depicts $\chi^2$ as a function of
$m_b$ for both neutrino mass orderings.
This figure clearly reflects the feature mentioned above
since for values of $m_b$
below $1190\:\mbox{MeV}$, $\chi^{2}$ increases dramatically and lower values
of $m_b$ become strictly ruled out.

There also exists an upper bound on $m_b$ in Figure~\ref{fig_mb}
at about $1250\:\mbox{MeV}$, which is located \emph{below} the central
input value of $m_\tau$ at $1292\:\mbox{MeV}$. In contrast to the
normal neutrino mass spectrum, however, the inverted spectrum seems to
prefer values for $m_b$ near its lower bound,
as can be read off from Figure~\ref{fig_mb}.

In summary,
our scenario imposes rather rigid constraints on $m_b$
and favors $b-\tau$ unification.
This feature is apparently caused by the $SO(10)$ relation~(\ref{Md-Ml})
between the mass matrices of charged-leptons and
down-quarks, which differ only by a factor of $-3$ in the off-diagonal
matrix elements.

\section{Conclusions}
\label{concl}

In this paper we have presented an attempt to combine an $SO(10)$ GUT
with the family symmetry $A_4$. We have considered renormalizable
Yukawa interactions, therefore, the choice of scalar $SO(10)$-plets
for fermion mass matrices is
confined to $\ten$, $\oht$ and $\hts$.
The three fermion families are accommodated in an
$A_4$ triplet. For fitting purposes we use the
fermion masses at the GUT scale evolved by the renormalization group
equations of the MSSM.
As a further important prerequisite we assume
that the VEVs occurring in the scalar
$SO(10)$ multiplets can be freely chosen for the purpose of fitting fermion
masses and mixings. Our investigation consists of two steps---for
the details see Section~\ref{sec-model}.

In the first step we have considered only the down-quark and
charged-lepton mass matrices. We have assigned all possible
$A_4$ representations to the $\ten$, $\oht$ and $\hts$ and checked, if
the down-quark and charged-lepton masses can correctly be
reproduced. In this way we have identified a unique successful scenario given
by the $\ten$ in $\ones$ of $A_4$ and the
$\hts$ in the $\mathbf{3}$ of $A_4$.
This is a non-trivial result, because we use only six masses for
probing mass matrices constructed with more than six parameters.
The mass matrices~(\ref{Md-Ml}) of the successful scenario~(\ref{case})
reflect the $SO(10) \times A_4$ structure: The 10-plets contribute a
general diagonal and the 126-plets a general off-diagonal matrix to
the mass matrices.

In the second step we have assumed type~II dominance
in the seesaw mechanism generating light neutrino masses.
Since only the 126-plets contribute to the neutrino mass matrix, we
obtain the Zee--Wolfenstein mass matrix.

The scenario gives an excellent
fit to all known data on fermion masses and mixings and shows,
therefore, the compatibility of the family group $A_4$ with $SO(10)$
GUTs. In summary, we have found the following features:
\begin{itemize}
\item
All mass matrices are symmetric.
\item
In the charged-fermion sector, as an effect of $SO(10) \times A_4$,
the building blocks of the mass matrices are general diagonal and
off-diagonal matrices, generated by the VEVs of
scalar 10-plets and 126-plets, respectively.
\item
$\mnu$ is given by the Zee--Wolfenstein matrix with its
definite predictions for the neutrino masses derived from
Eq.~(\ref{traceless}).
\item
The scenario can equally well accommodate normal and inverted neutrino
mass spectra.
\item
The lepton mixing angles of the Zee--Wolfenstein mass matrix which are
in disagreement with the data are corrected by contributions to
the PMNS matrix from $M_\ell$.
\item
There are definite predictions for $\meff$ for both spectra.
\item
Our scenario gives $b$--$\tau$ Yukawa unification.
\end{itemize}
On the negative side we note that our scenario does not feature
tri-bimaximal lepton mixing, however, some minor constraints on the lepton
mixing angles exist---see Sections~\ref{sec-normal} and
\ref{sec-inverted}. Moreover, up-quark masses and the CKM
matrix are completely free and can thus be adapted to the data without
imposing any restrictions on the parameters of the mass matrices
$M_d$, $M_\ell$ and $\mnu$.

\vspace{5mm}

\noindent
\textbf{Acknowledgments:} We thank L.~Lavoura for valuable suggestions
and reading the manuscript.


\newpage

\begin{thebibliography}{99}

\bibitem{fritzsch}
H. Fritzsch, P. Minkowski,
Ann. Phys. 93 (1975) 193.

\bibitem{seesaw}
P. Minkowski,
Phys. Lett. B 67 (1977) 421; \\
T.Yanagida,
in \textit{Proceedings of the Workshop on Unified Theory
and Baryon Number in the Universe},
O. Sawata and A. Sugamoto eds.,
KEK report 79-18, Tsukuba, Japan, 1979; \\
S.L. Glashow,
in \textit{Quarks and Leptons,
Proceedings of the Advanced Study Institute
(Carg\`ese, Corsica, 1979)},
J.-L. Basdevant et al. eds.,
Plenum, New York, 1981; \\
M. Gell-Mann, P. Ramond, and R. Slansky,
in \textit{Supergravity},
D.Z. Freedman and F. van Nieuwenhuizen eds.,
North Holland, Amsterdam, 1979; \\
R.N. Mohapatra, G. Senjanovi\'c,
Phys. Rev. Lett. 44 (1980) 912.

\bibitem{typeII}
G. Lazarides, Q. Shafi, C. Wetterich,
Nucl. Phys. B 181 (1981) 287; \\
R.N. Mohapatra, G. Senjanovi\'c,
Phys. Rev. D 23 (1981) 165; \\
R.N. Mohapatra, P. Pal,
\textit{Massive Neutrinos in Physics and Astrophysics},
World Scientific, Singapore, 1991, p. 127.

\bibitem{seesaw-general}
J. Schechter, J.W.F. Valle,
Phys. Rev. D 22 (1980) 2227; \\
S.M. Bilenky, J. Ho\v{s}ek, S.T. Petcov,
Phys. Lett. B 94 (1980) 495; \\
I.Yu. Kobzarev, B.V. Martemyanov, L.B.~Okun, M.G. Shchepkin,
Yad. Phys. 32 (1980) 1590
[Sov. J. Nucl. Phys. 32 (1981) 823]; \\
J. Schechter, J.W.F. Valle,
Phys. Rev. D 25 (1982) 774.

\bibitem{MSGUT}
C.S. Aulakh, R.N. Mohapatra,
Phys. Rev. D 28 (1983) 217; \\
T.E. Clark, T.K. Kuo, N. Nakagawa,
Phys. Lett 115B (1982) 26; \\
K.S. Babu, R.N. Mohapatra,
Phys. Rev. Lett. 70 (1993) 2845
[hep-ph/9209215]; \\
C.S. Aulakh, B. Bajc, A. Melfo, G. Senjanovi\'c, F. Vissani,
Phys. Lett. B 588 (2004) 196
[hep-ph/0306242].

\bibitem{detailed}
K. Matsuda, Y. Koide, T. Fukuyama,
Phys. Rev. D 64 (2001) 053015
[hep-ph/0010026]; \\
K. Matsuda, Y. Koide, T. Fukuyama, H. Nishiura,
Phys. Rev. D 65 (2002) 033008 (Err. \textit{ibid.} D 65 (2002) 079904)
[hep-ph/0108202]; \\
T. Fukuyama, N. Okada,
JHEP 11 (2002) 011
[hep-ph/0205066]; \\
B. Bajc, G. Senjanovi\' c, F. Vissani,
Phys. Rev. Lett. 90 (2003) 051802
[hep-ph/0210207]; \\
H.S. Goh, R.N. Mohapatra, S.P. Ng,
Phys. Lett. B 570 (2003) 215
[hep-ph/0303055]; \\
H.S. Goh, R.N. Mohapatra, S.P. Ng,
Phys. Rev. D 68 (2003) 115008
[hep-ph/0308197]; \\
B. Bajc, G. Senjanovi\'c, F. Vissani,
Phys. Rev. D 70 (2004) 093002
[hep-ph/0402140]; \\
K.S. Babu, C. Macesanu,
Phys. Rev. D 72 (2005) 115003
[hep-ph/0505200].

\bibitem{schwetz}
S. Bertolini, T. Schwetz, M. Malinsk\'y,
Phys. Rev. D 73 (2006) 115012
[hep-ph/0605006].

\bibitem{small120}
K. Matsuda, T. Fukuyama, H. Nishiura,
Phys. Rev. D 61 (2000) 053001
[hep-ph/9906433]; \\
S. Bertolini, M. Frigerio, M. Malinsk\'y,
Phys. Rev. D 70 (2004) 095002
[hep-ph/0406117]; \\
S. Bertolini, M. Malinsk\'y,
Phys. Rev. D 72 (2005) 055021
[hep-ph/0504241].

\bibitem{aulakh06}
C.S. Aulakh,
hep-ph/0602132; \\
W. Grimus, H. K\"uhb\"ock,
Phys. Lett. B 643 (2006) 182
[hep-ph/0607197]; \\
W. Grimus, H. K\"uhb\"ock,
Eur. Phys. J. C 51 (2007) 721
[hep-ph/0612132]; \\
C.S. Aulakh,
hep-ph/0607252; \\
C.S. Aulakh,
talk presented at \textit{33rd International Conference on High Energy Physics
  (ICHEP06)}, Moscow, Russia, July 26--August 2, 2006,
hep-ph/0610097; \\
C.S. Aulakh, S.K. Garg,
hep-ph/0612021.

\bibitem{MSGUTout}
C.S. Aulakh,
expanded version of the plenary talks at the
\textit{Workshop Series on Theoretical High Energy Physics},
IIT Roorkee, Uttaranchal, India, March 16--20, 2005,
and at the \textit{8th European Meeting
``From the Planck Scale to the Electroweak Scale'' (PLANCK05)},
ICTP, Trieste, Italy, May 23--28, 2005,
hep-ph/0506291; \\
B. Bajc, A. Melfo, G. Senjanovi\' c, F. Vissani,
Phys. Lett. B 634 (2006) 272
[hep-ph/0511352]; \\
C.S. Aulakh, S.K. Garg,
Nucl. Phys. B 757 (2006) 47
[hep-ph/0512224].

\bibitem{nu-review}
M. Maltoni, T. Schwetz, M.A. T\'ortola, J.W.F. Valle,
New. J. Phys. 6 (2004) 122
[hep-ph/0405172]; \\
G.L. Fogli, E. Lisi, A. Marrone, A. Palazzo,
Prog. Part. Nucl. Phys. 57 (2006) 742
[hep-ph/0506083].

\bibitem{S4xSO10}
D.G. Lee, R.N. Mohapatra,
Phys. Lett. B 329 (1994) 463
[hep-ph/9403201]; \\
C. Hagedorn, M. Lindner, R.N. Mohapatra,
JHEP 0606 (2006) 042
[hep-ph/0602244]; \\
Yi Cai, Hai-Bo Yu,
Phys. Rev. D 74 (2006) 115005
[hep-ph/0608022].

\bibitem{morisi}
S. Morisi, M. Picariello, E. Torrente-Lujan,
Phys. Rev. D 75 (2007) 075017
[hep-ph/0702034].

\bibitem{A4}
E. Ma, G. Rajasekaran,
Phys. Rev. D 64 (2001) 113012
[hep-ph/0106291]; \\
K.S. Babu, E. Ma, J.W.F. Valle,
Phys. Lett. B 552 (2003) 207
[hep-ph/0206292]; \\
M. Hirsch, E. Ma, A. Villanova del Moral, J.W.F. Valle,
Phys. Rev. D 72 (2005) 091301 (Err. \textit{ibid.} D 72 (2005) 119904)
[hep-ph/0507148]; \\
G. Altarelli and F. Feruglio,
Nucl.Phys. B 720 (2005) 64
[hep-ph/0504165; \\
G. Altarelli and F. Feruglio,
Nucl.Phys. B 741 (2005) 215
[hep-ph/0512103]; \\
E. Ma,
Phys. Rev. D 73, 057304 (2006)
[hep-ph/0511133]; \\
E. Ma, H. Sawanaka, M. Tanimoto,
Phys. Lett. B 641 (2006) 301
[hep-ph/0606103]; \\
L. Lavoura, H. K\"uhb\"ock,
Mod. Phys. Lett. A 22 (2007) 181
[hep-ph/0610050]; \\
and references therein.

\bibitem{altarelli}
G. Altarelli,
Lectures given at the 61st Scottish Universities Summer School in
Physics, St. Andrews, Scottland, 8--23 August 2006,
hep-ph/0611117.

\bibitem{HPS}
P.F. Harrison, D.H. Perkins, W.G. Scott,
Phys. Lett. B 530 (2002) 167
[hep-ph/0202074].

\bibitem{suitable}
E. Ma,
Mod. Phys. Lett. A 21 (2006) 2931
[hep-ph/0607190].

\bibitem{zee-model}
A. Zee,
Phys. Lett. 93B (1980) 389;
\textit{ibid.} 161B (1985) 141.

\bibitem{wolfenstein}
L. Wolfenstein,
Nucl. Phys. B 175 (1980) 93.

\bibitem{zee-pred}
C. Jarlskog, M. Matsuda, S. Skadhauge and M. Tanimoto,
Phys. Lett. B 449 (1999) 240
[hep-ph/9812282]; \\
P.H. Frampton and S.L. Glashow,
Phys. Lett. B 461 (1999) 95
[hep-ph/9906375].

\bibitem{kummer}
W. Konetschny and W. Kummer,
Phys Lett. 70B (1977) 433; \\
T.P. Cheng and L.-F. Lee,
Phys. Rev. D22 (1980) 2860.

\bibitem{he-zee}
Xiao-Gang He and A. Zee,
Phys. Rev. D 68 (2003) 037302
[hep-ph/0302201].

\bibitem{schwetz1}
T.~Schwetz,
Phys. Scripta T 127 (2006) 1
[hep-ph/0606060].

\bibitem{balaji}
K.R.S. Balaji, W. Grimus, T. Schwetz,
Phys. Lett. B 508 (2001) 301
[hep-ph/0104035]; \\
Y. Koide,
Phys. Rev. D 64 (2001) 077301
[hep-ph/0104226].

\bibitem{bimaximal-deviation}
C. Giunti, M. Tanimoto,
Phys.Rev. D 66 (2002) 053013
[hep-ph/0207096]; \\
C. Giunti, M. Tanimoto,
Phys. Rev. D 66 (2002) 113006
[hep-ph/0209169]; \\
P.H. Frampton, S.T. Petcov, W. Rodejohann,
Nucl. Phys. B687 (2004) 31
[hep-ph/0401206]; \\
K.A. Hochmuth, S.T. Petcov, W. Rodejohann,
arXiv:0706.2975.

\bibitem{sakita}
R.N. Mohapatra, B. Sakita,
Phys. Rev. D 21 (1980) 1062.

\bibitem{slansky}
R. Slansky,
Phys. Rept. 79  (1981) 1.

\bibitem{downhill}
J.A. Nelder, R. Mead,
Comp. J. 7 (1965) 306; \\
W.H. Press, B.P. Flannery, S.A. Teukolsky, W.T. Vetterling,
\textit{Numerical recipes in C: The art of scientific computing},
Cambridge University Press, 1992.

\bibitem{doublet-triplet1}
S. Dimopoulos, F. Wilczek,
in: \textit{The Unity of the Fundamental Interactions},
Proceedings of the 19th Course of the International
School of Subnuclear Physics, Erice, Italy, 1981,
edited by A. Zichini (Plenum Press, New York, 1983) 237-249;\\
K.S. Babu, S.M. Barr,
Phys. Rev. D 48 (1993) 5354
[hep-ph/9306242].

\bibitem{doublet-triplet2}
K.S. Babu, I. Gogoladze, Z. Tavartkiladze,
Phys. Lett. B 650 (2007) 49
[hep-ph/0612315].

\bibitem{minimalfinetuning}
C.S. Aulakh, A. Girdar,
Int. J. Mod. Phys. A 20 (2005) 865
[hep-ph/0204097]; \\
B. Bajc, A. Melfo, G. Senjanovi\'c, F. Vissani,
Phys. Rev. D 70 (2004) 035007
[hep-ph/0402122].

\bibitem{das}
C.R. Das, M.K. Parida,
Eur. Phys. J. C 20 (2001) 121
[hep-ph/0010004].

\bibitem{RPP}
W.-M. Yao et al., \textit{Review of Particle Physics},
J. Phys. G 33 (2006) 1.

\bibitem{antusch}
S. Antusch, J. Kersten, M. Lindner, M. Ratz,
Nucl. Phys. B 674 (2003) 401
[hep-ph/0305273].

\bibitem{raffelt}
See for instance \\
S.~Hannestad and G.~Raffelt,
JCAP 04 (2004) 008
[hep-ph/0312154]; \\
{\O}.~Elgar{\o}y and O.~Lahav,
New J. Phys. 7 (2005) 61
[hep-ph/0412075]; \\
A.~Goobar, S.~Hannestad, E.~M\"ortsell and H. Tu,
JCAP 06 (2006) 019
[astro-ph/0602155]; \\
M.~Fukugita, K.~Ichikawa, M.~Kawasaki and O.~Lahav,
Phys. Rev. D~74 (2006) 027302
[astro-ph/0605362]; \\
and references therein.

\end{thebibliography}
\end{document}